\documentclass[10pt, conference, letterpaper]{IEEEtran} 
\IEEEoverridecommandlockouts

\usepackage{graphicx}
\usepackage{float}
\usepackage{amsmath, amsthm}
\usepackage{dsfont}
 
\usepackage[ruled,vlined]{algorithm2e}
\usepackage[font=small]{caption}
\usepackage{subcaption}
\usepackage[export]{adjustbox}
\usepackage{siunitx}
\usepackage{colortbl}
\usepackage{algorithmic}
\usepackage[thinlines]{easytable}
\usepackage{siunitx}
\usepackage[normalem]{ulem}
\usepackage{amsfonts}
\usepackage{multirow,verbatim}
\usepackage[normalem]{ulem}
\usepackage[dvipsnames]{xcolor}
\usepackage{url}
\usepackage{subcaption}
\usepackage{multicol}
\usepackage{cite}
\usepackage{comment}

\SetKwComment{Comment}{$\triangleright$\ }{}

\newcolumntype{b}{>{\columncolor{blue!10}}c}
\newcolumntype{y}{>{\columncolor{yellow!10}}c}
\newcolumntype{d}{>{\columncolor{red!7}}c}

\newtheorem{observation}{Observation}

\SetKwInput{KwInput}{Input}                
\SetKwInput{KwOutput}{Output}              
\usepackage{booktabs,makecell,multirow,hhline,threeparttable} 
\usepackage{xcolor,colortbl,pgf}
\definecolor{Color1}{RGB}{240, 240, 240}
\usepackage{makecell}
\begin{document}

\title{Experimental Demonstration of Over the Air Federated Learning for Cellular Networks}
\author{
		\IEEEauthorblockN{
  Suyash Pradhan \IEEEauthorrefmark{1},
        Asil Koc\IEEEauthorrefmark{2},
        Kubra Alemdar\IEEEauthorrefmark{3},
        Mohamed Amine Arfaoui\IEEEauthorrefmark{2},
        Philip Pietraski \IEEEauthorrefmark{2},\\
        Francois Periard \IEEEauthorrefmark{2},
        Guodong Zhang \IEEEauthorrefmark{2}, 
        Mario Hudon \IEEEauthorrefmark{2}, and
        Kaushik Chowdhury\IEEEauthorrefmark{1}
       }
\IEEEauthorblockA{\IEEEauthorrefmark{1}Wireless Networking and Communications Group, The University of Texas at Austin, 
\IEEEauthorrefmark{2}InterDigital Inc. and \\
\IEEEauthorrefmark{3}Institute for the Wireless Internet of Things, Northeastern University \\
}
}
\maketitle

\begin{abstract}

Over-the-air federated learning (OTA-FL) offers an exciting new direction over classical FL by averaging model weights using the physics of analog signal propagation. Since each participant broadcasts its model weights concurrently in time and frequency, this paradigm conserves communication bandwidth and model upload latency. Despite its potential, there is no prior large-scale demonstration on a real-world experimental platform. This paper proves for the first time that OTA-FL can be deployed in a cellular network setting within the constraints of a 5G-compliant waveform. To achieve this, we identify challenges caused by multi-path fading effects, thermal noise at the radio devices, and maintaining highly precise synchronization across multiple clients to perform coherent OTA combining.
To address these challenges, we propose a unified framework for real-time channel estimation, model weight to OFDM symbol mapping and dual-layer synchronization interface to perform OTA model training. 
We experimentally validate OTA-FL using two relevant applications - Channel Estimation and Object Classification, at a large-scale on ORBIT Testbed and a portable setup respectively, along with analyzing the benefits from an operator's perspective. Under specific experimental conditions, OTA-FL achieves equivalent model performance, supplemented with $43\times$ improvement in spectrum utilization and $7 \times$ improvement in energy efficiency over classical FL when considering $5$ nodes.

\end{abstract}

\vspace*{-3pt}
\section{Introduction}
\label{sec:intro}
\vspace*{-7pt}

Federated Learning (FL) is a collaborative learning approach, where models are trained iteratively, with several training epochs completed locally on edge devices and then the model is averaged at a centralized server to obtain a global model for dissemination~\cite{PFL, fl_mom}. However, in typical IoT applications, these recurring and massive model updates from hundreds or even thousands of nodes will quickly saturate the uplink, leading to lengthy convergence time and network resource consumption. This paper studies the feasibility of a newly emerging paradigm called over-the-air FL (OTA-FL) that uses the physics of signal propagation to combine signals in flight. Despite some early theoretical works and initial proof-of-concepts, OTA-FL has never been demonstrated in practice at a large-scale with realistic operational conditions of a cellular network, primarily due to the simplifying assumptions of the environment. This paper shows this concept experimentally while addressing many unsolved challenges in realizing OTA-FL using an actual 5G-compliant waveform. 

\begin{figure}
    \centering
    \includegraphics[width=0.3\textwidth]{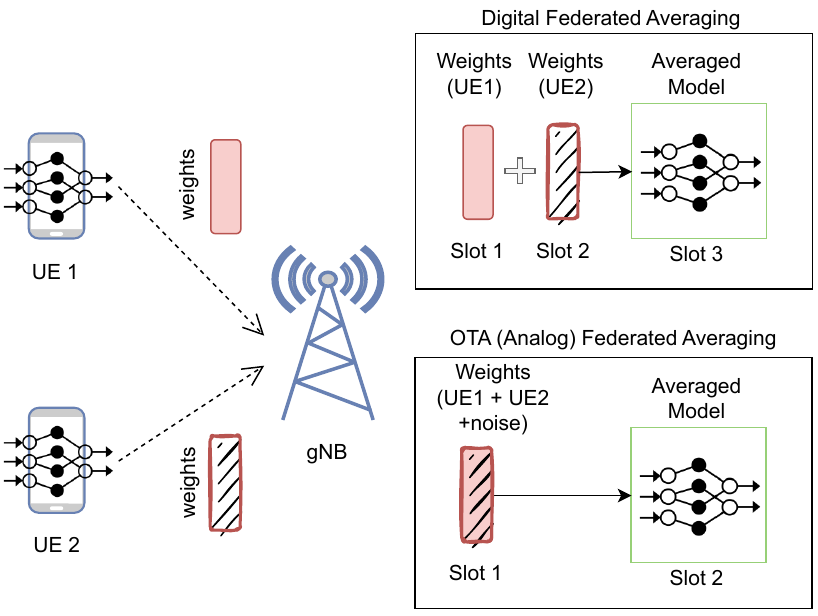}
    \caption{
    Comparison of Digital FL (top) requiring dedicated time slots for sending weights and digital averaging, while OTA-FL (bottom) combines signals over-the-air in an analog fashion using time slots required by a single UE. However, OTA-FL's performance is influenced by receiver noise, channel effects, and precise synchronization, which are investigated in this paper.}
    \label{fig:Figure1}
    \vspace{-5mm}
\end{figure}

\vspace{-1mm}
\subsection{How to conduct Federated Learning by effectively utilizing the wireless spectrum?}

Collaborating the roles of computation and communication to meet users demands has been an active area of research \cite{airfc}, \cite{icompcomm}.
Along this direction, a novel concept of over the air computation has emerged in the wireless communication community. The key idea is to utilize the signal superposition of analog signals to compute a couple of fundamental mathematical functions over the wireless channel. It offers substantial benefits in resource utilization, specifically by minimizing communication delay and enhancing bandwidth efficiency for edge computing scenarios. This methodology aligns well with the practical requirements of FL, in terms of enabling communication-efficient model aggregation. Multiple clients (UEs in this scenario) can simultaneously access the same time-frequency resources to transmit weights in analog format, which are aggregated over the air. The server (gNB in our case) can then scale these aggregated weights to perform federated averaging (FedAvg) \cite{FL_1} efficiently, utilizing the wireless channel as a computational resource.

\vspace{-1mm}
\subsection{Scenario of Interest and Challenges} In Fig \ref{fig:Figure1}, we consider two UEs performing local model training and sending weights for averaging to obtain the global model. The upper part highlights transmission using digital communication, wherein each device requires an independent time slot $t_1$ and $t_2$ for UE1 and UE2, respectively, to transfer their model parameters to the server, with the time slots increasing exponentially for each additional UE. On the other hand, the bottom figure highlights the OTA-FL paradigm, where all the weights are transmitted in an analog format over the same time slot $t_1$, regardless of the number of participants, performing the aggregation over the wireless channel itself. Thus, the OTA-FL methodology possesses benefits in terms of bandwidth, latency and efficient compute. However, OTA-FL using analog signals poses a couple of technical challenges:

\noindent$\bullet$ \textbf{Imperfect Channel Estimation}: Signals transmitted with limited power and combining in the air will be influenced by range-dependent path losses.  We need to design a suitable pre-equalization technique to maintain amplitude-phase alignment. 


\noindent$\bullet$ \textbf{Thermal Noise during Signal Aggregation}: Given the limited transmit power budget at UEs and the presence of channel fading, the additive noise at the gNB becomes significant, leading to fluctuations in FL model training. A strategy is needed to mitigate this additive noise across training rounds, ensuring smoother and consistent training cycles.

\noindent$\bullet$ \textbf{Stringent Synchronization Requirements}: To perform coherent aggregation, the signals need to be transmitted precisely at the same time, with the synchronization order expected to reach the sub-$\mu s$ level, keeping it below the sample duration. The nature of this experimentation necessitates testing on a real-world testbed with actual system constraints.

\vspace{-1mm}
\subsection{Contributions}




\noindent$\bullet$
We introduce the first-of-its-kind real-world, model-agnostic OTA-FL demonstration, featuring both a large-scale implementation using the ORBIT Testbed \cite{ORBIT}, with multiple UEs positioned 20m from the gNB to emulate typical UE-gNB separation in cellular networks, and a portable setup, with both utilizing USRP X310 SDRs (Software Defined Radios).

\noindent$\bullet$ 
We design a channel inversion precoding and power control scheme to overcome the wireless effects taking into account the impact of real hardware impairments, while mapping updates to an OFDM waveform. We implement a two-level synchronization interface, utilizing Precision Time Protocol (PTP) \cite{PTP} for accurate hardware timestamping at the Host level, along with an Octoclock distribution system to synchronize the radio devices, to ensure synchronization at the symbol duration level aligning with 5-G NR standards.

\noindent$\bullet$ We evaluate OTA-FL using two relevant applications: channel estimation and object classification. Our analysis focuses on the influence of the different gradient distributions, along with insights to fine-tune hyperparameters for performing OTA model aggregation with real hardware. OTA-FL demonstrates comparable model performance to its digital counterpart, converging significantly faster in terms of absolute time, due to common network resources. We quantify the large-scale benefits in terms of spectrum conservation and energy efficiency.

\noindent$\bullet$ We release the software implementation for OTA-FL with USRP devices at \cite{genesys_ota_fl_testbed}, thereby providing researchers with the tools to replicate, and apply their independent precoding strategies and models for OTA-FL with minimal modifications.
\section{Related Work}
\label{sec:related_work}
\vspace*{-3pt}


 We highlight some key works in the OTA-FL space to understand the current research landscape. In \cite{BAA},  Zhu \textit{et al.} introduce analog aggregation for FL in a low-latency cellular scenario, under the assumption of perfect CSI. Furthermore, in \cite{CHARLES}, Mao \textit{et al.} focus on developing adaptive power control and control local updates based on channel conditions emulated using the small-scale Rayleigh fading model.
In \cite{OTA_FL_Perf_CSI}, Yang \textit{et al.} focuses on solving joint device selection and beamforming to perform FL under the assumption of perfect CSI at the clients. In \cite{OTA_LR}, Xu \textit{et al} conduct a work on performing learning rate optimization for OTA-FL in a dynamic fashion in order to adapt to the uplink fading channel effects. Over the two works \cite{COTAF}, \cite{COTAF_J}, Sery \textit{et al} propose a time-varying precoding scheme to mitigate the noise effects under ideal channel conditions and extension to fading channels and validate it using a simulation study with heterogeneous data. There have been some initial prototypes in prior literature. In \cite{OTA_FL_Prototype}, Guo \textit{et al.} propose an equalization technique to address timing offsets and validate it using $2$ IoT sensor clients on a toy task with a shallow $2$ layer neural network. Miao \textit{et al.} \cite{1-bit-OTA-FL} introduce 1-bit quantization as an effective digital method of performing OTA-FL. In a series of studies \cite{FSK_MV_OFDM, FSK_MV_Bias, FSK-MV}, Sahin \textit{et al.} introduce encoding techniques to implement this scheme, with validation carried out in an indoor environment using SDRs at a 5-meter distance. However, this work is confined to the SignSGD optimizer, which may not always be optimal.

\textbf{Novelty over state-of-the-art:} 
OTA-FL has primarily been explored in simulations, with a few proof-of-concepts limited to specific models and quantization schemes, which may compromise model performance in certain cases. Additionally, these approaches do not address challenges such as long-range communication, common in cellular network scenarios. In this work, we bridge this gap by conducting OTA-FL on a large-scale real-world testbed, using a precoding strategy suited for analog circuitry and mapping model updates from any type of DL model to a 5G-compatible OFDM waveform. We explore two use cases from different domains, highlighting the model-independent nature of our approach, and evaluate ML performance, spectrum efficiency, and energy gains, demonstrating its broader applicability across diverse cellular network applications.

\vspace{0mm}


\vspace{2mm}
\vspace*{-3pt}
\section{OTA FL System Model}
\label{sec:framework}
\vspace*{-3pt}

\begin{figure}
    \centering
    \includegraphics[trim=0 0 0 0,clip,width=0.3\textwidth]{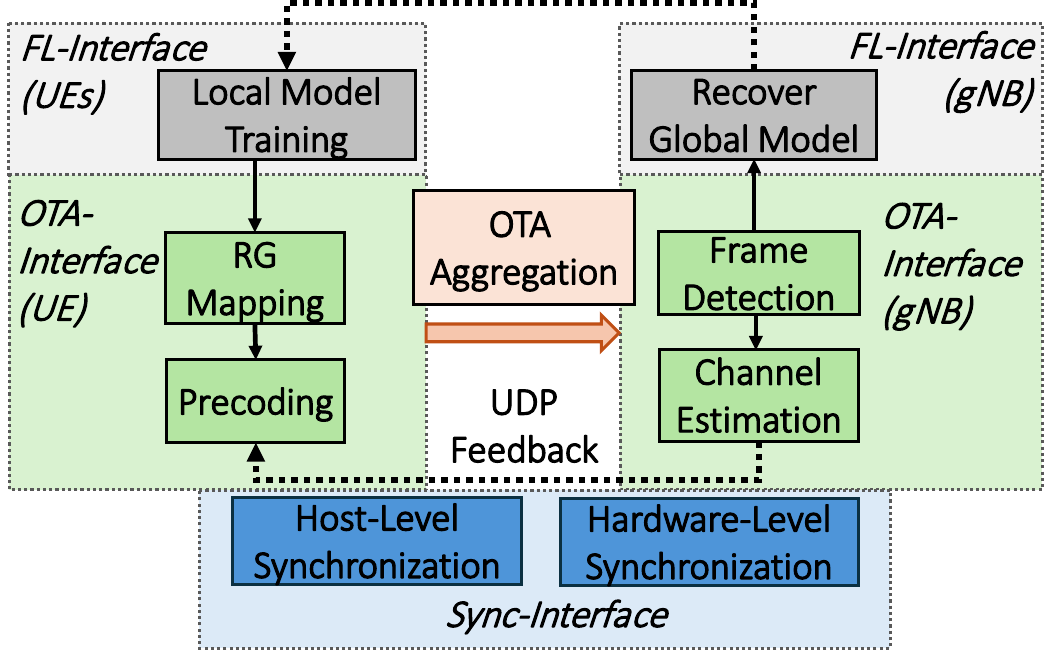}
    \caption{End-to-end framework for OTA-FL using USRP devices}
    \label{fig:framework}
\end{figure}

In this section, we describe our system model starting from the enhanced CSI estimation, RG mapping and precoding along with satisfying the transmit power constraints. The blocks of the end-to-end process are illustrated in Fig. \ref{fig:framework}.

\subsection {Local Model Training}

We consider $M$ participating UEs where $i$ denotes the UE number such that $i = {1, 2, ... M}$, each training their local models. For the initial communication round, each UE trains their local model and computes local parameters $\Theta_i$ after completing their $E$ local training epochs. 
 We need to design the ML model and fine-tune the hyperparameters to ensure that the computed updates remain within an acceptable range for the analog circuits (DAC, PA, Mixers). Simply scaling by the peak value could cause the average and smaller updates to become insignificant. Therefore, we propose adjusting the ML model hyperparameters to minimize the variance across the local model parameters during a communication round, thereby preventing gradients from vanishing or exploding to levels that are unacceptable for the transmit-side electronics and wireless communication. Additionally in OTA-FL, we suggest transmitting the difference between the received global model and the updated local model in subsequent rounds, rather than the weights themselves. This enables incremental stable learning despite the inherent noise, especially at longer distances or under poor channel conditions. The model update is computed as $\Delta \Theta_{i, r} = \Theta_{i, r}$ - $\Theta_{r - 1}^g$, which will help the model in learning gradually in the presence of noise - \(\Theta_{r}^g = \Theta_{r - 1}^g + \frac{1}{M}\sum_{i = 1}^M \Delta \Theta_{i, r}\)


\subsection{Channel Estimation}

We conduct pilot-aided channel estimation to compensate for the wireless effects which will be encountered during OTA combining. We generate a set of QAM symbols denoted by 
$X_{m, n}$ at symbol $m$ and sub-carrier $n$, where $m$ and $n$ can be scalars or vectors depending on pilot density. These symbols experience the wireless downlink channel $H_{m, n}$. 
The received pilots can be expressed mathematically as follows:
\begin{equation}
    Y_{m, n} = (X_{m, n} * H_{m, n}) + \mathcal{N_D}
\end{equation}

where $\mathcal{N_D} \sim \mathcal{N}(0,\,\sigma_D^{2})$  represents the thermal noise experienced at the receiver.
The receiver apply LS estimator, dividing the received pilots $Y_{m, n}$ with the known pilot symbols $X_{m, n}$ to get the channel estimates $\hat{H}_{m, n}$ at pilot locations.

\vspace{-2mm}
\begin{equation}
\vspace{-1mm}
\hat{H}_{m, n} = (Y_{m, n} / X_{m, n}) 
\end{equation}

Thereafter, these pilot estimates $\hat{H}_{m, n}$ are interpolated over the entire RG such that $\hat{H}_D = I(\hat{H}_{m, n})$, where $I(x)$ is the interpolation function, to obtain entire estimated downlink channel $\hat{H_D}$. We measure the performance of the estimated downlink channel $\hat{H}_D$ w.r.t true channel $H_D$, for $N$ data samples, in terms of NMSE (Normalized Mean Square Error): \(= \frac{1}{N} \sum_{s = 1}^{N} \frac{|\hat{H}_{s, D} - H_{s, D}|^2}{|H_{s, D}|^2} \)

\subsection{Resource Grid Mapping}

 

Here, we consider each model has their respective model updates $U_{i, r}$ computed at their nodes. We independently scale the I and Q components of the symbols with the peak value across the model updates in that component. We denote the set of scaled model updates for a particular i$^{th}$ client using a set $U_{i, r}$ = ${\{u_{i, r}^1, u_{i, r}^2, .... u_{i, r}^P\}}$ where $P$ is the total number of model parameters. We take advantage of the fact that signals are sent as complex numbers in OFDM grids with in-phase and quadrature components. This enables us to encode a pair of weights into 1 symbol in the OFDM RG, thus saving half the resource space. Thus, the mapping of weights takes place in the following manner -$
w_{i, k} = u_{i, r}^{2k -1} + j u_{i}^{2k} $, where $k = \{1, 2, ...P/2\}$ Resource Elements (REs). 
To split model parameters into the frame slots, we consider a bandwidth of $B$ Hz with a sub-carrier spacing of $\delta$ Hz. Using this, we can obtain the total number of sub-carriers along the RG which can be computed as ($B$ / $\delta$). The number of symbols in a RG is denoted by $T$. With this, one RG comprises of [$T$ * ($B$ / $\delta$)] symbols. Finally, number of slots $\rho$ comes down to $\rho = \frac{P}{2 * T * (B / \delta)} $. The model updates in the OFDM slots are denoted by $W_i = \{w_{i, 1}, w_{i, 2} ... w_{i, P/2}\}$.

\subsection{Precoding the Parameters}
Each UE will utilize the channel estimates \( \hat{H}_i^f \) received as feedback to precode the signal for compensation of channel effects. Each device will follow a channel-inversion precoding scheme \cite{CHARLES} and transmit the precoded model updates as \( \frac{W_i}{\hat{H}_i^f} \). The goal is to ensure that the overall signal transmitted by the UE stays within  the maximum uplink power constraint \( P_{M}^U \). To achieve this, we scale the overall signal to ensure it does not exceed the peak power constraint. Considering the applied precoding scheme, we define a power control value \( \alpha \) as the transmitter gain and ensure that the average power constraint is met, ensuring both the signal value \( W_i \) and the estimated \( \hat{H}_i^f \) fall within the desired range.


\vspace{-2mm}
\begin{align}
      \left|\alpha *  \frac{W_i} {\hat{H}_{i}^f} \right|^2 &< P_{M}^U \quad
    \alpha < \sqrt{\frac{P_{M}^U}{ \left| \frac{W_i} {\hat{H}_{i}^f}\right|^2}}
\end{align}

 


\subsection {OTA Aggregation} Ultimately, the signals are transmitted over the air, with uplink channel $H_U$, using the same resources in time and frequency, expecting the experienced uplink channel and precoded vector to compensate each other. Thus, we compute the superposition of signals $S_i$ from the UEs. With minimal misalignment, we expect to approximately recover the aggregated signal as follows:
\vspace{0mm}
\begin{align}
Y^O &= 
\alpha * \sum_{i = 1}^{M} \frac{(H_U  * W_i)}{\hat{H}_i^f} + \mathcal{N_U}
 &\sim= \alpha * \sum_{i = 1}^{M} (W_i) + \mathcal{N_U}
\end{align}

\subsection {Post-Processing at gNB}
We obtain the original signals from the received slots, decoupling the precoding factor and scaling by 
$M$ to convert summation to averaging, such that -
$
\Delta \Theta_{r}^{g} = Y^O / (M * \alpha)$. 
We insert these averaged model updates to obtain global model parameters for the current round. Thereafter, we retrieve the global model parameters $\Theta_{ r}^{g}$ from the RG by interleaving the real and imaginary parameters as shown in Fig. \ref{fig:map}, which are sent via an error-free downlink channel back to the UEs to continue model training. 

\begin{figure}[t!]
    \centering
    \begin{subfigure}{0.9\linewidth}
    \centering
    \includegraphics[trim=0 0 0 0,clip,width=0.9\linewidth]{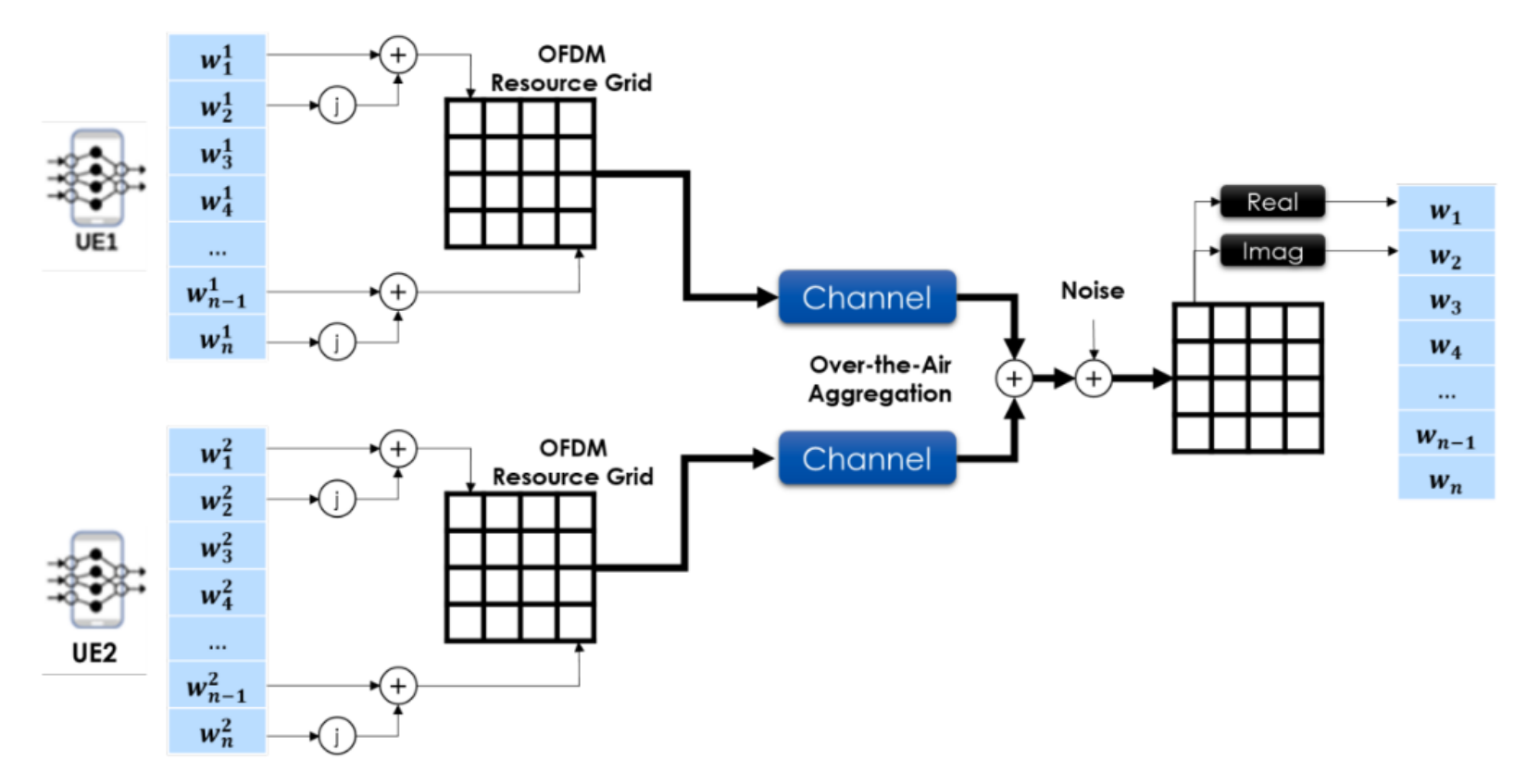}
    \caption{}\label{fig:map}
    \label{fig:map}
    \end{subfigure}\hspace{-10mm}
     \begin{subfigure}[b]{0.6\linewidth}
       \includegraphics[width=1.2\textwidth]{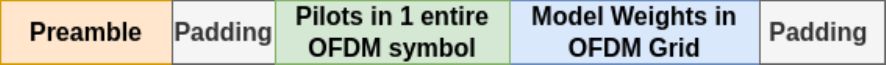}
         \caption{}
         \label{fig:Frame_Real}
     \end{subfigure}
    \vspace{0mm}
    \caption{a) Encoding Model Parameters to OFDM Frame ; b) frame structure used in real-world experiments}
   
    \end{figure}
     \vspace{0mm}

\section{Real World OTA-FL Framework}

This section details the real-world OTA-FL experimental setup and software stack. While we conduct experiments for the Channel Estimation use case at a large scale using the ORBIT (Open-Access Research Testbed for Next-Generation Wireless Networks) testbed \cite{ORBIT}, and Object Classification using a portable version, the overall software orchestration framework remains consistent, as described in this section. We emulate UEs and a gNB setup using multiple USRP X310 SDRs. The software framework consists of three main components: the OTA Aggregation Interface, the Synchronization Interface, and the FL Interface (Model Training).

\subsection{OTA Aggregation Interface}

\noindent$\bullet$ \textbf{Frame Detection:} The overall frame structure is shown in Fig. \ref{fig:Frame_Real}. 
Our custom OFDM Frame consists of three components - preamble used for frame detection, pilots used for channel estimation and the main payload which carries the model updates in the OFDM waveform. 
This preamble is designed for synchronizing the starting sample of the frame and includes Gold Sequences (GS) \cite{Gold}, known for their distinct correlation properties. This helps us provide precise sample-level synchronization prior to each OTA aggregation.

\noindent$\bullet$ \textbf{Channel Estimation:}
The second field of the frame comprises pilot signals, akin to the CSI-RS used in 5-G NR. In this configuration, we dedicate 1 OFDM symbol for transmitting known QPSK symbols. This approach is suitable because the nodes remain stationary during real-world experimentation, resulting in gradual changes in a flat-fading channel. For simplicity, we adopt simplex communication where both pilot and data transmissions occur in the uplink. The UEs transmit pilot signals to the gNB, where the LS estimator calculates the channel vector across individual sub-carriers. The CSI feedback is then sent over a UDP socket to the UEs, enabling them to apply channel-inversion precoding to the model parameters before transmission.



\subsection{Synchronization Interface}

A real-time implementation of OTA-FL requires multi-level synchronization, encompassing both compute nodes and communication devices (SDRs). We categorize our synchronization approach into two levels:

\noindent$\bullet$ \textbf{Host Level Synchronization:} Here, we manage synchronization for the nodes emulating UEs. To achieve this, we implement the Precision Time Protocol (PTP), a Computer Network protocol designed to synchronize clocks uniformly across computers. PTP operates over Ethernet networks and utilizes a master-slave architecture to distribute precise timing information. PTP is specified in the IEEE 1588 standard \cite{PTP} and provides robust timing accuracy at a sub-$\mu s$ level.
PTP gains an advantage from hardware timestamping, where timing information is directly extracted from the register in the Network Interface Card (NIC) of the computing node, contingent upon the capabilities of the device. 
In our setup, we configure the computer serving as the gNB to the Master, while all client computers act as slaves receiving clock synchronization from the gNB. We implement Linux tools, including ptp4l, to synchronize all the hardware clocks and phc2sys to propagate individual hardware timestamps to the system level. 
This setup ensures that all UEs maintain synchronized time, facilitating coordinated operations. 

\noindent$\bullet$ \textbf{Communication System Synchronization:} We rely on the ORBIT infrastructure for SDR synchronization. Each mini-rack has an Octoclock-G providing PPS and 10MHz clock signals to all SDRs, which are all synchronized to a master reference clock. Equal-length cables and symmetric topology ensure minimal phase and time deviations among SDRs.

\subsection{FL Interface and Orchestration}

The local models for different clients are trained using Tensorflow \cite{tensorflow} for a certain pre-determined number of local epochs. We extract the model weights from the different model layers and store them in a single array. We scale the parameters to satisfy the power constraints and transfer them to the GNURadio SDR Framework for mapping these updates to IQ samples spanning across time-frequency resources and perform IFFT as a part of OFDM modulation. We provide the PTP timestamp to the UHD driver function to ensure the multiple host computers controlling the SDRs can start transmitting exactly at the same time and at the same rate, keeping up to precise synchronization both in time and frequency. This way the signals overlap and can perform coherent OTA combining during propagation. At the gNB unit, we detect the frame start and perform FFT as a part of OFDM demodulation to get the combined model updates in the form of time-frequency grid. We conduct some post-processing operations to recover the averaged weights in an array, which are inserted into the global model layers and monitor the NMSE of OTA aggregated weights compared to the ones digitally averaged after each communication round. We iterate this process until convergence according to the approach discussed in Sec. \ref{sec:framework}.

\vspace*{-3pt}
\section{OTA-FL Use Cases}
\label{sec:simulation}

\subsection{Channel Estimation}

\begin{figure}[t!]
    \centering
    \begin{subfigure}{0.55\linewidth}
    \centering
        \includegraphics[trim=5 0 0 0, height=2cm,,width=1\linewidth]{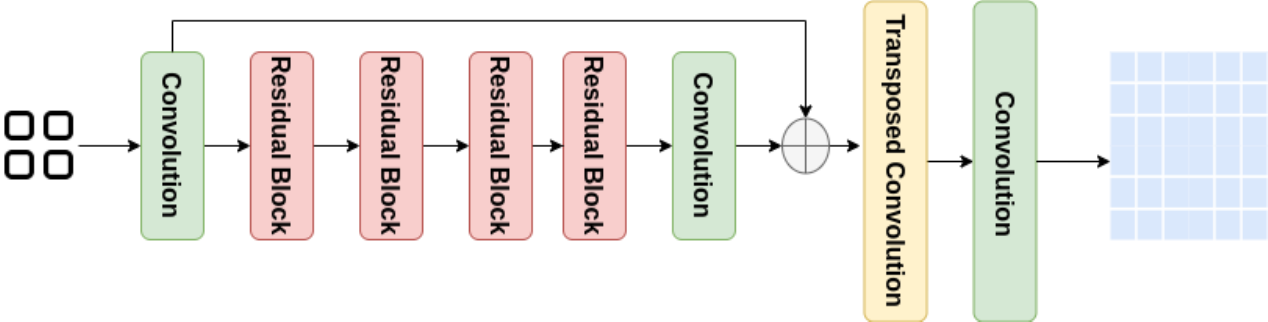}
        \caption{ReEsnet Channel Estimator}
        \label{fig:CSI_Model}
    \end{subfigure}%
    \hspace{1mm}
    \begin{subfigure}{0.4\linewidth}
    \centering
        \includegraphics[trim=0 0 0 84, clip, height = 2.5 cm, width=1\linewidth]{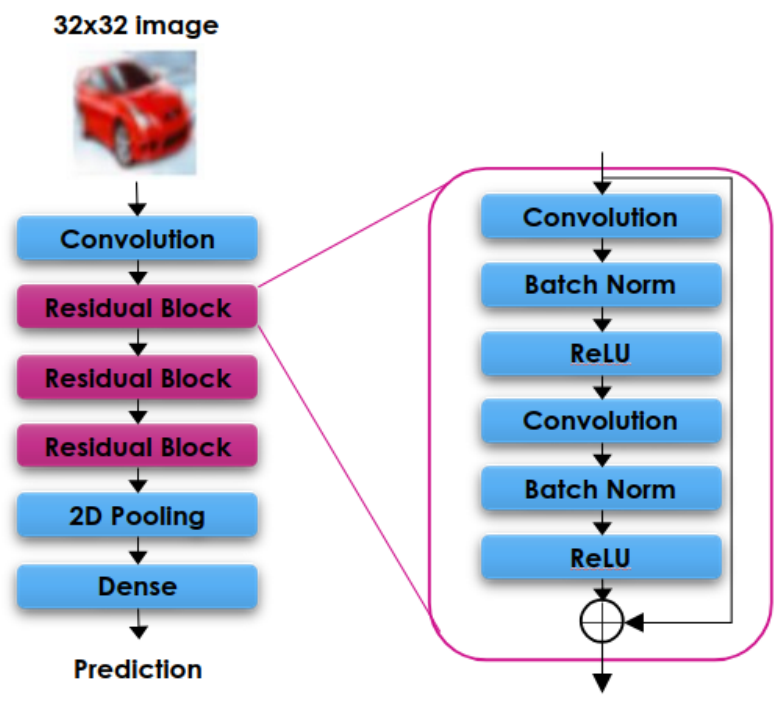}
        \caption{ResNet-8 Object Classifier}
        \label{fig:Image_Model}
    \end{subfigure}
     \begin{subfigure}{0.47\linewidth}
    \centering
    \vspace{3mm}
        \includegraphics[trim=5 0 0 0, clip, height=2cm,,width=1\linewidth]{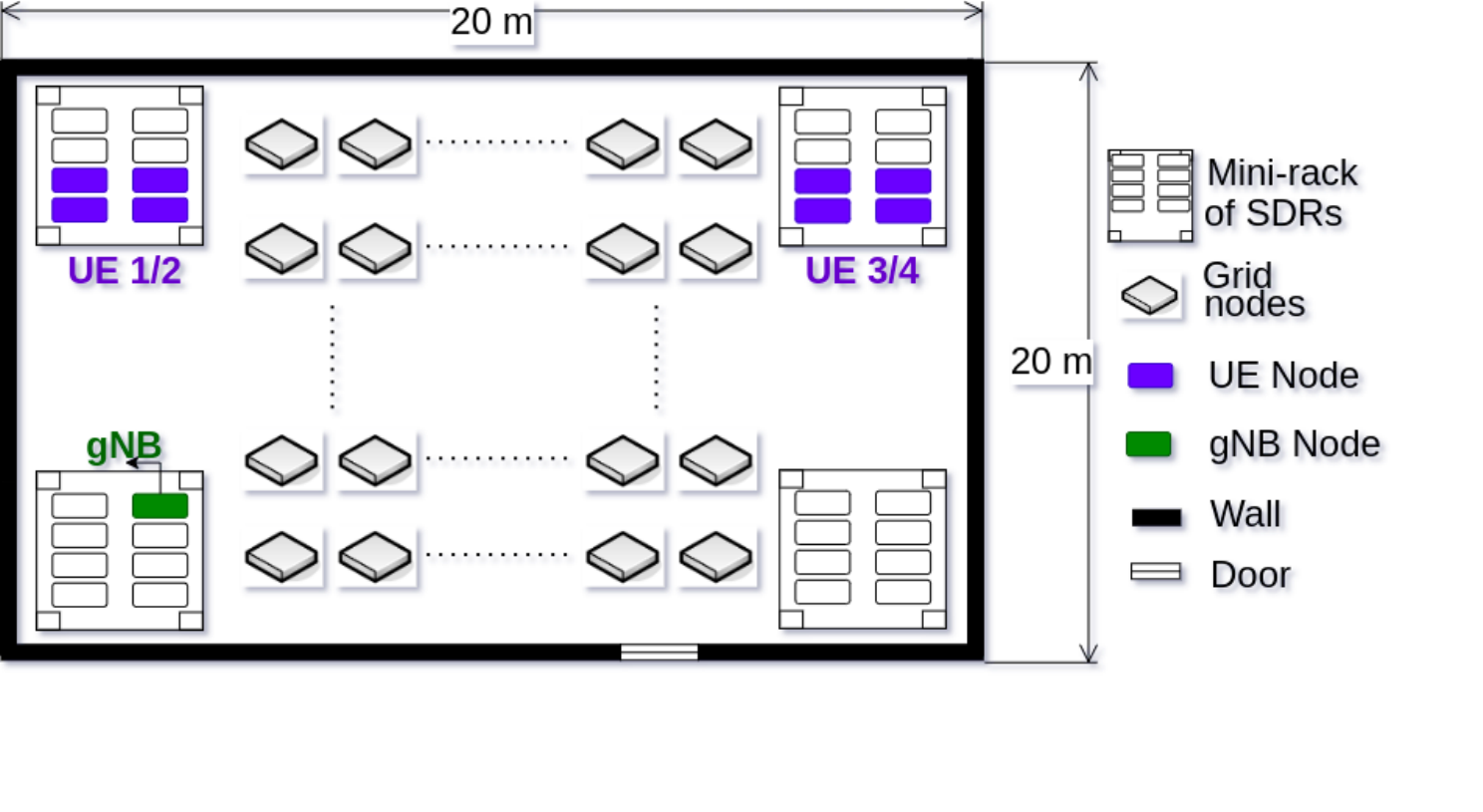}
        \caption{Large-scale ORBIT Testbed}
        \label{fig:ORBIT_Placement}
    \end{subfigure}%
    \hspace{1mm}
    \begin{subfigure}{0.47\linewidth}
    \centering
        \includegraphics[trim=60 0 60 170, clip, height = 2 cm, width=1\linewidth]{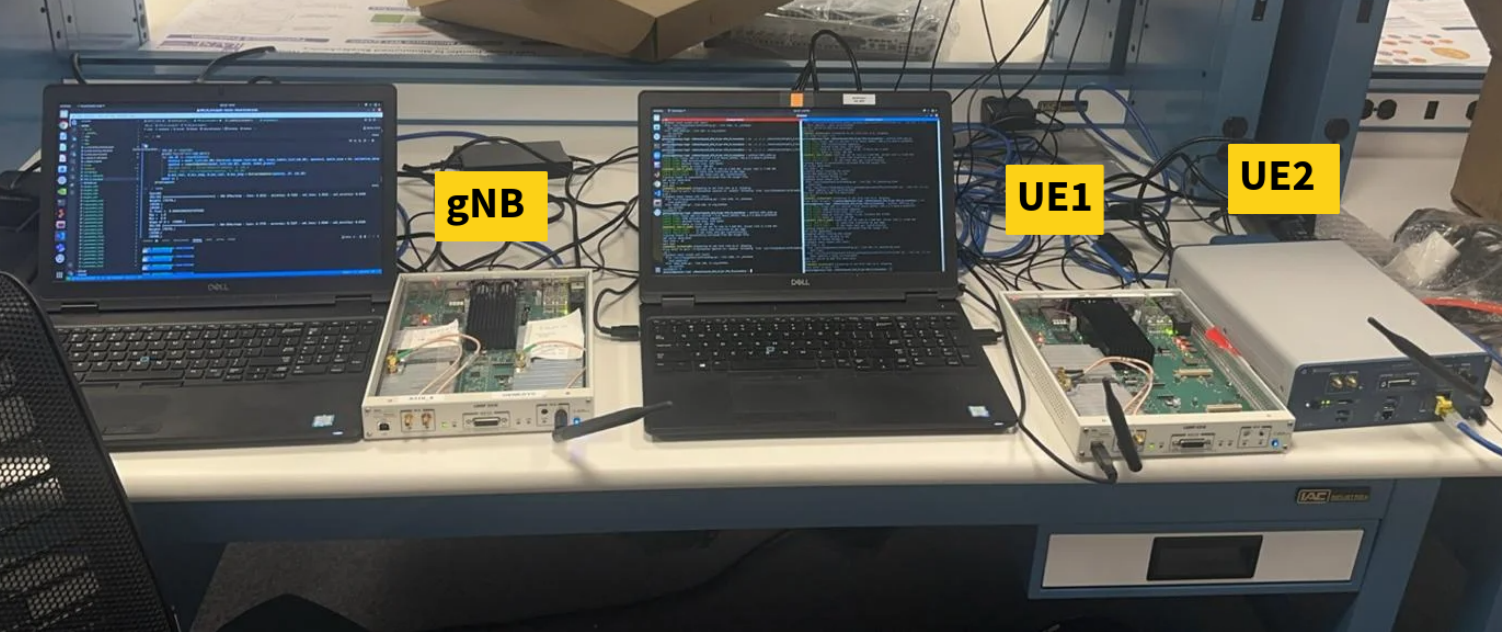}
        \caption{Portable Testbed Setup}
        \label{fig:Portable}
    \end{subfigure}
    \vspace{0mm}
    \caption{Model Architectures and Testbeds}
    \vspace{-4mm}
    \label{fig:time_sync}
    \vspace{-1mm}
    \end{figure}

We implement a Channel Estimation use-case using the large-scale version at the ORBIT Testbed, with $5$ UEs placed at $20m$ distance from the gNB as shown in Fig \ref{fig:ORBIT_Placement}.

\vspace{0mm} 
    \noindent$\bullet$ \textbf{Dataset Generation:} We create a simulation environment using NVIDIA Sionna \cite{sionna}, incorporating a UMi (Urban Micro Cell) channel model with a 100 m cell radius. The cellular range is divided into 5 clusters (10-20 m, 30-40 m, 50-60 m, 70-80 m, 90-100 m), with the gNB at the origin. Path loss and shadow fading effects introduce heterogeneity between the clusters. The transmit power at the gNB is 41 dBm for a $3.5$ GHz carrier and $10$ MHz bandwidth per UE, while the noise variance is set to $-174$ dBm/Hz \cite{3GPP}. The UE mobility range is $3-33.33$ m/s, simulating various outdoor conditions. The orientation and LOS/NLOS settings are randomly assigned. With a $10$ MHz bandwidth and $30$ kHz subcarrier spacing, the Resource Grid (RG) consists of $288$ subcarriers (calculated as $10$ MHz / $30$ kHz). There are $24$ Resource Blocks (RBs), each with $12$ subcarriers and $14$ symbols. A single CSI-RS pilot is inserted at symbol $6$ and subcarrier $6$ in each RB. We generate $20,000$ channel samples for each cluster, resulting in a total of $100,000$ CSI samples across the 5 clusters. The dataset is divided into $80\%$ training, $10\%$ validation, and $10\%$ global test data. 


     \noindent$\bullet$ \textbf{Model Architecture:} We select ReEsNet \cite{ReEsNet} as our lightweight ML model for its compact size and effective performance, with modifications to input a single pilot symbol and produce a larger $14 \times 288$ size RG, resulting in $71,666$ model parameters. The architecture is shown in Fig. \ref{fig:CSI_Model}. The model includes 4 local residual layers, a global residual skip connection for gradient propagation, and a Transposed Convolution layer for learnable upsampling to interpolate channel estimates over the entire RG. We consider a learning rate of $0.0001$ and batch size of $32$ with an Adam Optimizer \cite{Adam}.


\subsection{Object Classification}

We leverage the portable setup to conduct OTA-FL on the Object Classification use-case, as shown in Fig. \ref{fig:Portable}.

 \noindent$\bullet$ \textbf{Dataset Generation:} We use the CIFAR-10 dataset, which consists of 60,000 color images, each of size 32x32 pixels, spread across 10 distinct classes. The dataset is divided into 50,000 training images and 10,000 test images, providing a well-balanced sample for both training and evaluation purposes. We divide the training dataset images equally between the participating nodes.  Each client receives a random number of samples from each class, leading to varying class distributions among the clients.

  \noindent$\bullet$ \textbf{Model Architecture:} Here, we use a ResNet-8 architecture as shown in Fig \ref{fig:Image_Model}, a smaller variant of the ResNet family, consisting of 8 layers.  
  This architecture has proved to be particularly effective in CIFAR-10 object classification due to its ability to learn complex features while maintaining a relatively simple and computationally efficient structure. We perform the local FL training with a learning rate of $0.0008$, batch size of $64$ and $1$ local epoch using Adam Optimizer \cite{Adam}.

\vspace*{-3pt}
\section{Performance Evaluation}
\label{sec:experiments}
\vspace*{-3pt}

In this section, we focus on experimentally validating our methods for OTA-FL training on the real testbed. We analyze the synchronization accuracy achieved, the convergence trend with respect to absolute time and quantify the benefits in spectrum utilization.

\begin{figure}[t!]
    \centering
    \begin{subfigure}{0.49\linewidth}
    \centering
        \includegraphics[trim=0 0 0 0 ,width=1\linewidth]{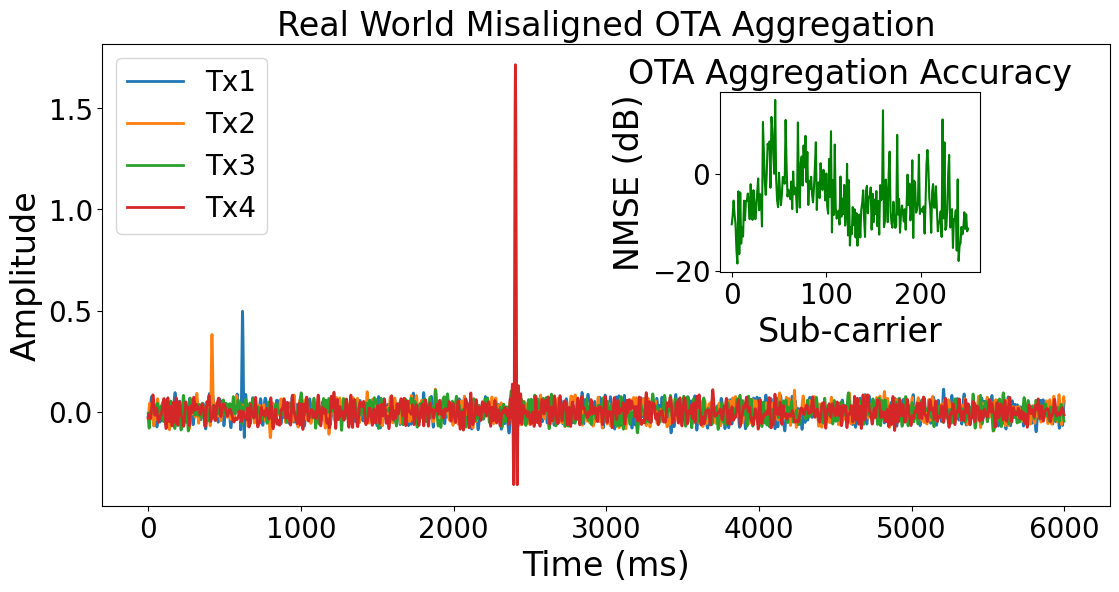}
        \caption{}
        \label{fig:Sync_Imperfect}
    \end{subfigure}%
    \begin{subfigure}{0.48\linewidth}
    \centering
        \includegraphics[trim=0 0 0 0,width=1\linewidth]{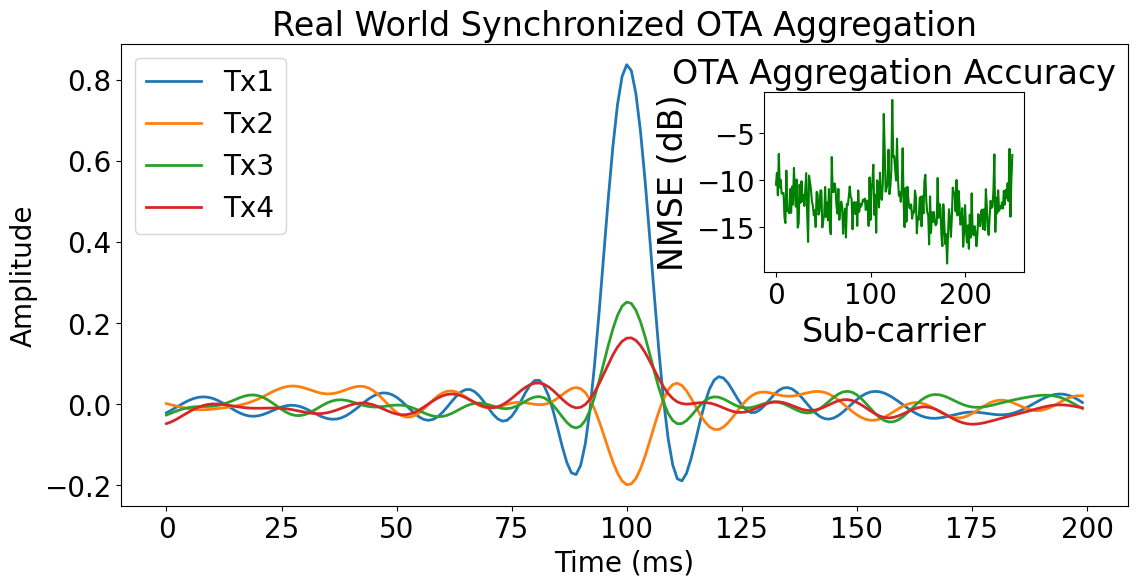}
        \caption{}
        \label{fig:Sync_Perfect}
    \end{subfigure}
    \vspace{-2mm}
    \caption{Synchronization accuracy for real-time OTA-FL: a) Correlation inaccuracy due to host-level misalignment; b) Aligned correlation peaks after PTP synchronization}
    \vspace{-4mm}
    \label{fig:time_sync}
    \vspace{-1mm}
    \end{figure}

\begin{figure}[t!]
\centering
\begin{subfigure}
[b]{0.23\textwidth}
    \includegraphics[trim=0 0 0 -2,clip,width=1\linewidth]{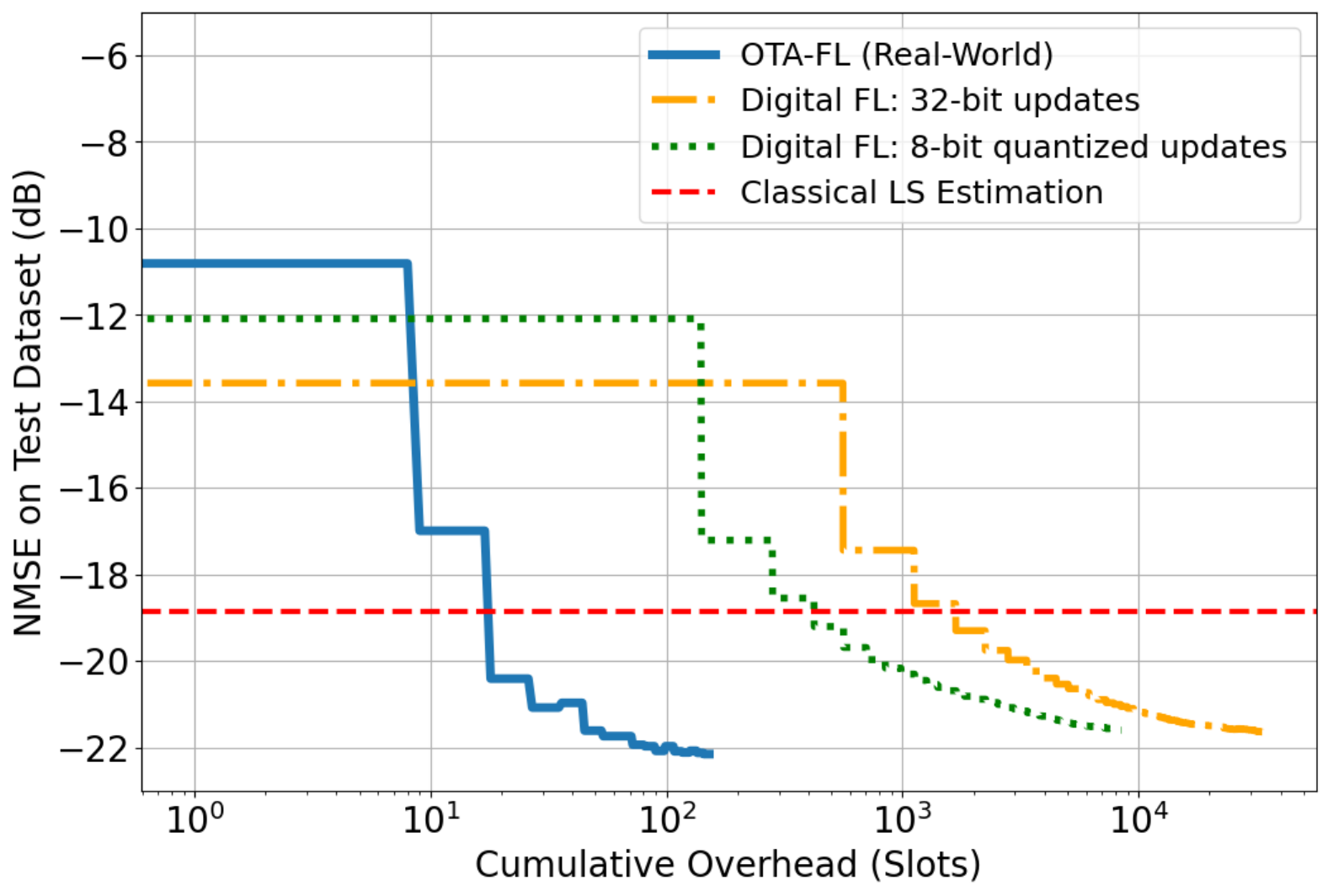}
   \centering  
   \subcaption{Channel Estimation}\label{fig:conv_csi}
\end{subfigure}\begin{subfigure}
[b]{0.23\textwidth}
    \includegraphics[trim=0 0 0 -2,clip,width=1\linewidth]{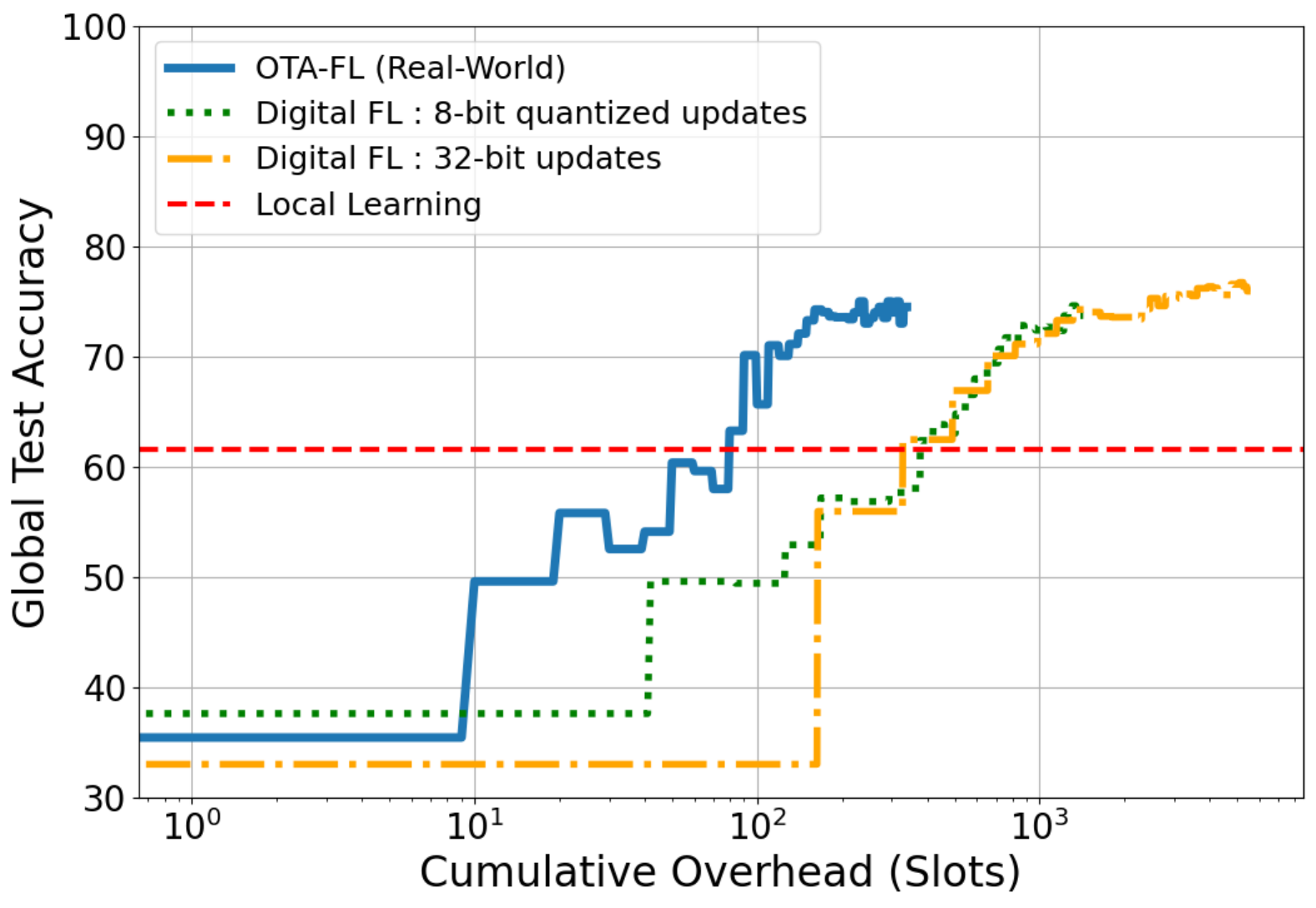}
   \centering
    \subcaption{Object Classification}\label{fig:conv_image}
\end{subfigure}

\vspace{0mm}
\caption{Convergence Analysis with respect to time slots consumed} \label{fig:spectrum}
\vspace{0mm}
\end{figure}

\vspace{0mm}

\begin{figure*}
\begin{subfigure}[b]{0.33\textwidth}
    \includegraphics[trim=0 0 0 0,clip,width=1\linewidth]{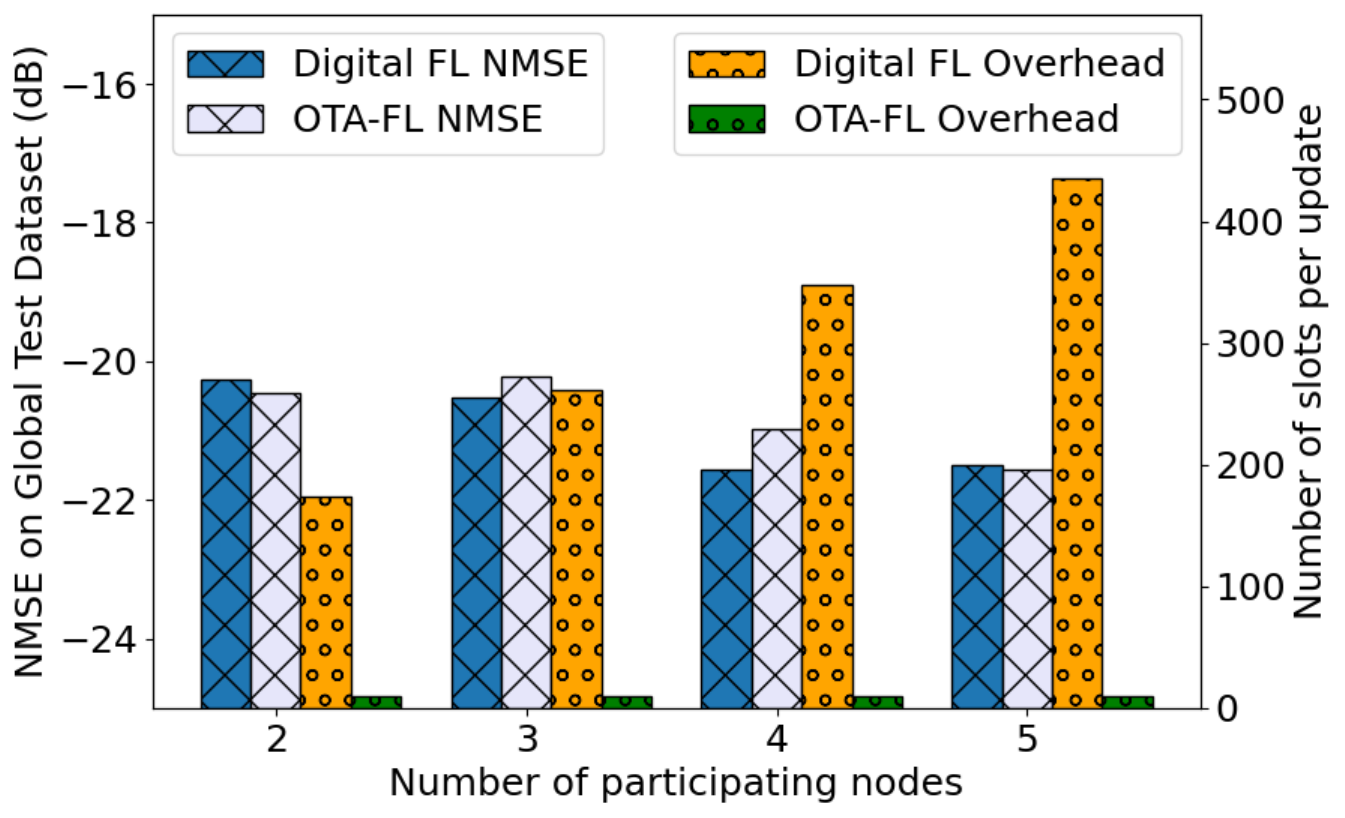}
    \centering
    \subcaption{}\label{fig:real_test}
\end{subfigure}
  \begin{subfigure}[b]{0.3\textwidth}
    \includegraphics[trim=0 0 0 0,clip,width=1\linewidth]{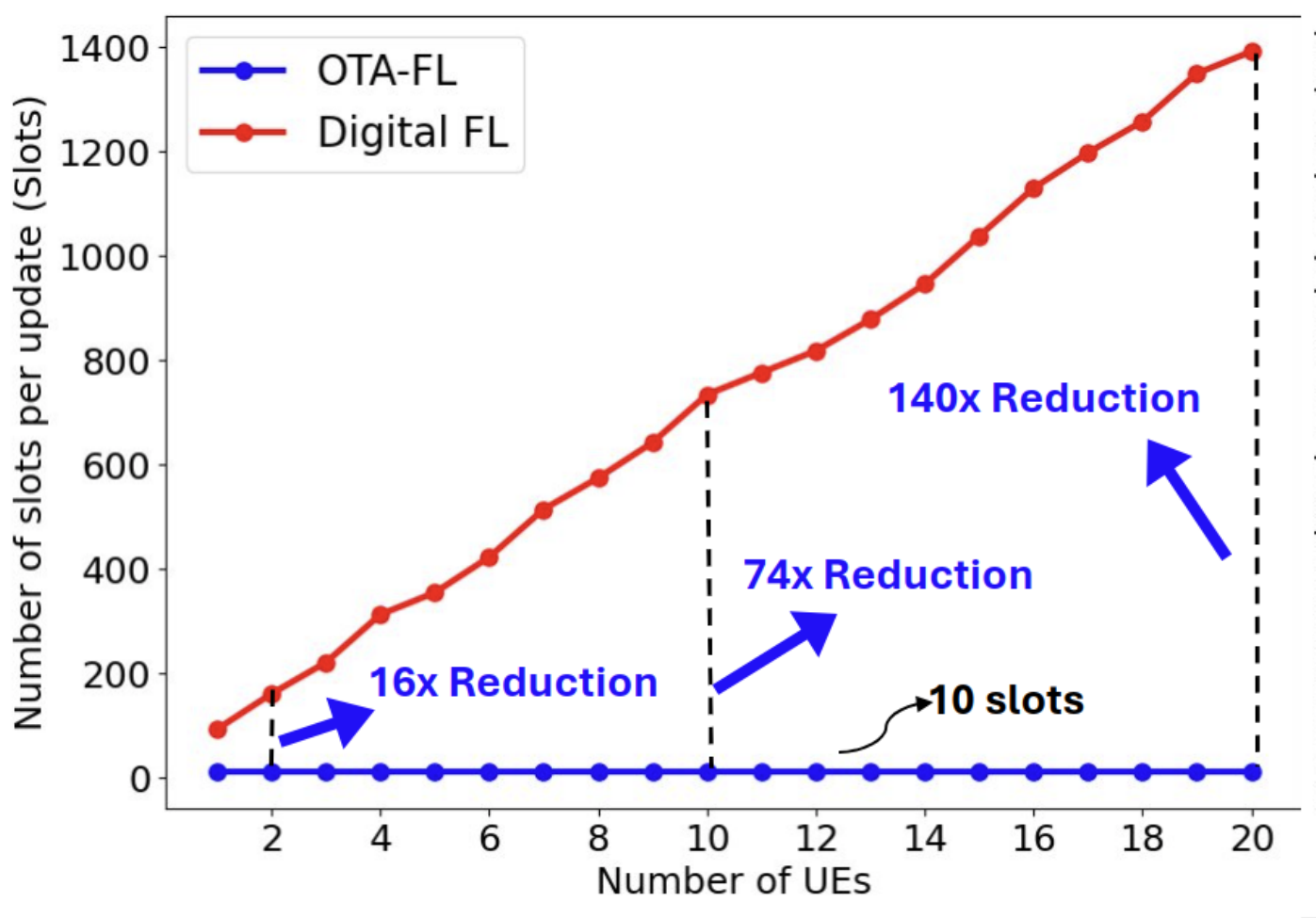} 
    \centering
    \subcaption{}\label{fig:overhead}\label{fig:overhead}
  \end{subfigure}
\begin{subfigure}
[b]{0.3\textwidth}
    \includegraphics[trim=0 0 0 0,width=1\linewidth]{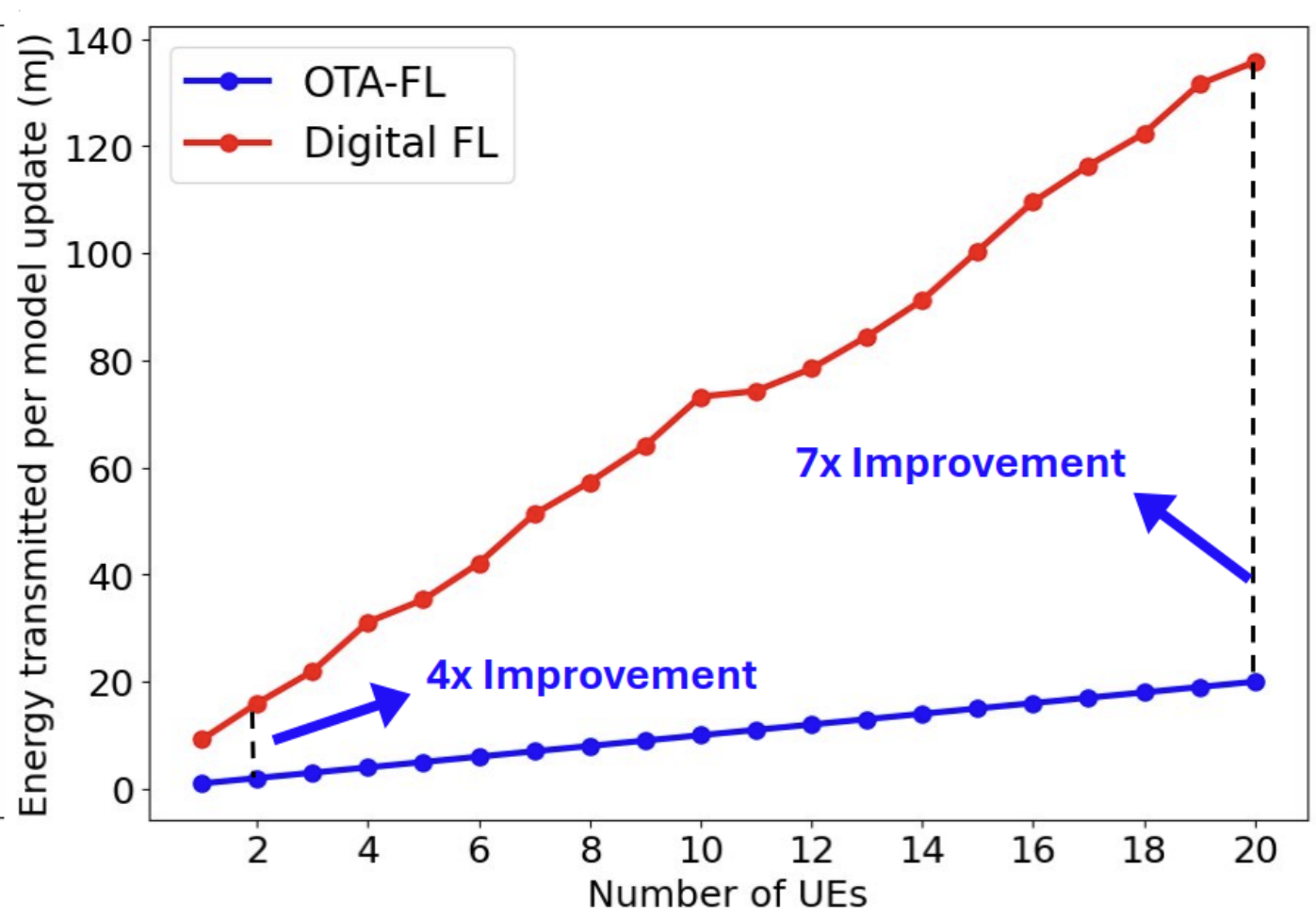}
   \centering
    \subcaption{}\label{fig:energy}
\end{subfigure}

\vspace{-2mm}
\caption{a) Model Performance vs Overhead on ORBIT Testbed; b) Communication Overhead;  c) Energy Efficiency} \label{fig:cons_corr}
\vspace{-5mm}
\end{figure*}



\subsection{Synchronization Accuracy}
To begin, we aim to evaluate the level of synchronization achieved after implementing the synchronization strategies and how well it aligns with the symbol duration as per the 5G NR standards. Each UE is allocated a bandwidth of $3.84$ MHz and $256$ sub-carriers, which results in a sub-carrier spacing of $3.84$ MHz / $256$ = $15$ kHz. The PHY setup follows numerology 0 of the 5G-NR frame structure, where the symbol duration is $1 / 15000$ = $66$ $\mu$s. In Fig. \ref{fig:Sync_Imperfect}, we present the correlation peaks of GS from different UEs, which are influenced by misalignment when PTP is disabled causing inadequate synchronization, along with the corresponding NMSE of OTA aggregation comparing the averaged OTA weights with the perfectly averaged weights. Then, in Fig. \ref{fig:Sync_Perfect}, we show a scenario where perfect peak alignment is achieved following PTP synchronization across all UEs, resulting in improved OTA averaging and reduced error variance. We can conclude that achieving frame synchronization at a sub-$\mu$s level throughout the entire processing and communication chain enables practical implementation of OTA averaging, with each symbol consuming $66 \mu s$, aligning with 5G NR numerology 0.


\subsection{Convergence Analysis and Mitigation of Thermal Noise}

This sub-section focuses on comparing OTA-FL convergence with state-of-the-art FL methods, with an emphasis on reducing communication overhead. Our baseline methods include Digital FL with full precision 32-bit and quantized 8-bit model updates. In Fig. \ref{fig:conv_csi}, we observe the performance for the Channel Estimation use-case in terms of NMSE in dB, whereas in Fig. \ref{fig:conv_image} we monitor the accuracy of object classification on the respective global test datasets. In both cases, we plot the convergence with respect to number of time slots required until convergence as a representative measure of absolute time and spectrum consumed until the models converge. Digital FL with full precision updates consumes maximum number of communication resources, to yield optimal performance, whereas OTA-FL showcases equivalent performance in both cases along with quicker convergence saving significant network resources. 
Using 8-bit quantized updates, we observe that there is a certain reduction in communication overhead. However, OTA-FL can outperform any model compression method due to the consumption of communication resources required for a single client, irrespective of the number of actual participants as well as pure analog processing which eliminates the overhead due to modulation and channel coding. In the Channel Estimation use-case, since the computed gradients connect to lower-order channel gain values, we observe that the gradient magnitude plays a role in the learning process and coupling OTA-FL with 1-bit quantization schemes, like state-of-art works, can deter performance. Nevertheless, we were able to obtain stable gradients with controlled variance suitable for OTA-FL, for this application. However, in object classification, our gradients had a higher variance initially, which required careful hyperparameter tuning, towards lower learning rates and higher batch sizes, to ensure the gradients are stable with lower variance and are suitable for analog transmissions. This provides us an insight about developing joing communication-learning design approaches for OTA-FL as an interesting research direction. 

\subsection{Scalability and Benefits of OTA-FL}
\textbf{Real World Experimentatal Analysis: }In this subsection, we analyze the reduction in communication overhead with increasing participants, using the Channel Estimation use-case. The ORBIT testbed consists of radio devices in massive MIMO mini-racks, with the gNB SDR at one corner and UEs' SDRs $20m$ away. For this experiment, we vary the number of participants from 2 to 5 to assess OTA-FL scalability under real hardware and over-the-air conditions. As shown in Fig. \ref{fig:real_test}, model performance remains consistent for both OTA-FL and Digital FL as the number of participants increases. The communication overhead, measured as time slots required for model updates, with each slot comprising complex $14$ symbols and $256$ subcarriers with spacing $15KHz$, a slot duration is $1ms$, with the ML model compirising of $71666$ parameters. Overhead remains constant for OTA-FL at 10 slots per round, while Digital FL overhead increases with each additional participant, calculated as $\sum_{k=1}^{M} \frac{(P \times b)}{m_k \times K}$, where $b$ is bit precision, $P$ is the number of model parameters, $K$ is the number of REs per slot, $M$ is participants, and $m_k$ is spectral efficiency. Using Sionna's 5G PUSCH simulation with a $20m$ UE-gNB separation, spectral efficiency is 7.4063, with a modulation order of 8 and coding rate of 0.92. For 32-bit updates, the total time for a Digital FL update is $86.396$ slots per participant. Thus, for $5$ UEs we observe Digital FL consumes $435$ slots, whereas OTA-FL still requires only $10$ slots, yielding almost a $43 \times$ reduction.

\textbf{Effective Spectrum Utilization: }After establishing the overhead reduction for $5$ UEs through real-world experimentation, we extrapolate this result uptil $20$ UEs, as shown in Fig. \ref{fig:overhead}. We can observe that for a $20$ UE case, there is almost a $140 \times$ reduction in time slots, highlighting the huge benefits in spectrum utilization for telecom operators, when deploying collaborative AI applications if they adopt the OTA-FL methodology.

\textbf{Energy Efficiency: } OTA-FL significantly reduces the number of time slots required for transmitting model updates in the uplink channel, leading to improved energy efficiency. Energy consumption is a function of power and time, so reducing the duration of transmissions directly decreases energy usage at both the UEs and gNB. For simplicity, we assume all UEs transmit at 20 dBm (an avergae 3 dBm less than maximum) in both OTA-FL and Digital FL. OTA-FL requires only 10 slots, while Digital FL requires 87 slots. As shown in Fig. \ref{fig:energy}, the energy required for a model update in OTA-FL increases marginally, whereas Digital FL exhibits a steeper rise. With 2 UEs, we observe a $4 \times$ improvement, and with 20 UEs, the improvement is $7 \times$. This is primarily due to the advantages of analog signal processing, which avoids complex modulation and coding schemes. The gNB also benefits from this lower energy consumption. Advances in O-RAN will enable dynamic sleep modes at the gNB, allowing it to power down during the saved duration or balance load. \cite{energy} 

\vspace{-1.5mm}

\vspace*{-3pt}
\vspace{2mm}
\section{Conclusions}
\label{sec:conclusions}
\vspace*{-2pt}

In this paper, we demonstrate for the first time OTA-FL at a large scale under real-world experimental settings. We address challenges including peak transmit power, multi-path fading, receiver thermal noise and stringent synchronization across multiple UEs to enable OTA combining. 
We develop an orchestration framework to conduct the entire process right from local model training to precoding, mapping the updates to an OFDM waveform and aggregating updates OTA maintaining synchronization both in time and frequency for coherent aggregation using USRP devices. We will be releasing the entire codebase to the research community for further investigations with independent model architectures and precoding mechanisms. Our evaluation highlights essential synchronization accuracies required to achieve OTA-FL in practice, aligned with the latest 5-G cellular standards. We also assess scalability across varying distances and participants. Our results show comparable performance to Digital FL along with significant benefits in terms of spectrum  and energy efficiency. As a future research direction, we plan to leverage the portable setup to conduct experiments introducing mobile UE scenarios, better emulating real channel conditions. 

\section*{Acknowledgement}
{This is a collaborative work with InterDigital Inc. We acknowledge their support and constructive feedback to shape this work, along with funding from the U.S. National Science Foundation (grant CNS-2112471).} 

\bibliographystyle{IEEEtran}  
\bibliography{references}     

\end{document}